\documentclass[final,journal,onecolumn]{IEEEtran}
\usepackage{cite}
\usepackage{graphicx}
\usepackage{color}
\usepackage{pifont}
\begin{document}

\title{The NIM Inertial Mass Measurement Project}
\author{Shisong Li, Zhonghua~Zhang, Qing He, Zhengkun Li, Wei Zhao, Bing Han,
        and~Yunfeng Lu
\thanks{Shisong Li is with the Department
of Electrical Engineering, Tsinghua University, Beijing 100084, China and National Institute of Metrology, Beijing 100013, China. E-mail: leeshisong@sina.com.}
\thanks{Zhonghua Zhang, Qing He, Zhengkun Li, Bing Han and Yunfeng Lu are with National Institute of Metrology, Beijing 100029, China. Wei Zhao is with the Department
of Electrical Engineering, Tsinghua University, Beijing 100084, China. }
\thanks{Manuscript received August 18, 2014; revised October 22, 2014.}}

\markboth{IEEE Trans. Instrum. Meas. (CPEM2014)}%
{IEEE Trans. Instrum. Meas. (CPEM2014)}

\maketitle

\begin{abstract}
An inertial mass measurement project, which is expected to precisely measure the Planck constant, $h$, for possible comparisons with known gravitational mass measurement projects, e.g., the watt balance and the Avogadro project, is being carried out at the National Institute of Metrology, China. The principle, apparatus, and experimental investigations of the inertial mass measurement are presented. The prototype of the experiment and the Planck constant with relative uncertainty of several parts in $10^{4}$ have been achieved for principle testing.
\end{abstract}

\begin{IEEEkeywords}
the Planck constant, inertial mass measurement, capacitance measurement, mechanical oscillator.
\end{IEEEkeywords}

\IEEEpeerreviewmaketitle

\section{Introduction}
\IEEEPARstart{R}{ecently}, the precision measurement of the Planck constant $h$ with its relevance for possible new definitions of the kilogram has become a topic of great concerns in metrology \cite{1,2,3}. Two strategies, the watt balance by comparing electrical power and mechanical power \cite{4} and counting atoms with determining the Avogadro constant $N_{A}$ (known as the Avogadro project) \cite{5}, are now being pursued by National Metrology Institutes (NMIs) towards the determination of the Planck constant with relative uncertainty of several parts in $10^{8}$. As is known, the gravitational mass is measured for both the watt balance and the Avogadro project. Therefore, any method realizing precision inertial mass measurement would be informational, which also meets the CIPM-2005 recommendation that 'further encourage NMIs to pursue national funding to support continued relevant research in order to facilitate the changes suggested here and improve our knowledge of the relevant fundamental constants, with a view to further improvement in the International System of Units' \cite{6}.

A simple idea for inertial mass measurement is to employ a mechanical oscillator. For example, a pendulum method was tried in Istituto Nazionale di Ricerca Metrologica (INRIM, Italy) for deriving the inertial mass from electrical quantities \cite{7}. However, the accuracy is limited because of the difficulty in measurement for the equivalent mass center of the pendulum. Further, the quasi-elastic electrostatic oscillation method, which eliminates the mass center measurement by a beam-balance oscillator, was proposed in \cite{8} for precision inertial mass measurement at National Institute of Metrology (NIM, China). The Planck constant is determined by comparing period changes of a beam-balance oscillator when weighing different masses and applying different quasi-elastic electrostatic forces. The elastic electrostatic force is generated by a twin-Kelvin-capacitor system with easy alignment procedure. Compared with the voltage balance experiment \cite{10}, the applied dc voltage is only several kilovolt (typical 1000V), thus the uncertainty due to the resistance divider \cite{11} is reduced. Besides, the measurement is insensitive to the air buoyancy as well as the local gravity acceleration $g$.

 The NIM inertial mass measurement project started in 2010 and a primary experimental apparatus has been built for principle testing. In this paper, we report the status of the inertial mass project with presenting several experimental investigations and an initial measurement of the Planck constant.

\section{Principle}
\begin{figure}[!t]
\centering
\includegraphics[width=0.55\columnwidth]{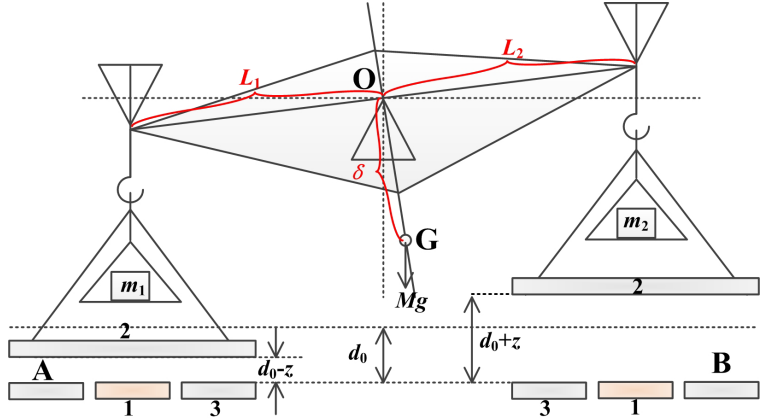}
\caption{Schematic diagram of the quasi-elastic electrostatic oscillation method. O is the central knife and G is the mass center of the balance beam (OG$=\delta$). $L_1$ and $L_2$ are left and right beam lengths. $m_1$ and $m_2$ are weighing masses. The movable electrodes of the twin Kelvin capacitor system (A and B) moves vertically in oscillation. Electrode 1 is the high potential terminal while electrodes 2 and 3 are grounded. $d_0$ is the distance between electrodes 1 and 2 at balancing. }
\label{Principle}
\end{figure}
Details of the principle and theoretical analysis for inertial mass determination based on quasi-elastic electrostatic oscillation method have been reported in \cite{8}, and a brief review is presented here. The differential equation for the oscillation system as shown in Fig.\ref{Principle} is expressed as (\ref{eq:diff}) with small oscillation amplitude around the balancing position.
\begin{equation}
(\beta_{1}+2m)\frac{d^{2}\theta}{dt^{2}}+(\beta_{2}-kU^{2})\theta=0.
\label{eq:diff}
\end{equation}
In (\ref{eq:diff}), $m=m_{1}=m_{2}$ is the weighing test mass; $\theta$ is the pivot angle; $U$ is the applied dc voltage between electrodes 1 and 2 (3); $\beta_{1}$ and $\beta_{2}$ are mechanical factors that defined as
\begin{equation}
\beta_1=\frac{J_0}{L^2}+2m_0,~~
\beta_2=\frac{M_0g\delta}{L^2},
\end{equation}
where $J_0$ denotes the moment of inertia of the balance beam, $m_0$
the mass of the suspension (except for the test mass), $L=L_1=L_2$ the beam length of the balance oscillator and $M_0$ the mass of the balance beam. Here we define $z$ as the vertical upward displacement of electrode 2B with respect to its equilibrium position, and $k$, the second order capacitance coefficient along the vertical direction $z$, can be written as
\begin{equation}
k=\frac{\partial^2 (C_{12A}+C_{12B}+C_{13A}+C_{12B})}{\partial z^2},
\end{equation}
where $C_{12A}$, $C_{12B}$, $C_{13A}$, and $C_{13B}$ is the capacitance between electrodes 1 and 2, 1 and 3 for capacitor A and B respectively. The periodic solution for (\ref{eq:diff}) is written as
\begin{equation}
T=2\pi\sqrt{\frac{\beta_{1}+2m}{\beta_{2}-kU^{2}}}.
\end{equation}
In order to obtain the SI-1990 electrical ratio $\gamma$ and the Planck constant $h$, the SI value for $\beta_1$ and the electrical value for $\beta_{2}$ should be known. Here the substitution method is applied to solve these two values. The first step is to make $U=0$, and the oscillation periods $T_1$ and $T_2$ are measured with different test masses $m=m_1$ and $m=m_2$ respectively. Then $\beta_{1}$ is solved as
\begin{equation}
\beta_{1}=\frac{2(m_{1}T_{2}^{2}-m_{2}T_{1}^{2})}{T_{1}^{2}-T_{2}^{2}}.
\label{eq.b1}
\end{equation}
It can be seen from (\ref{eq.b1}) that $\beta_1$ is determined in SI value. Similarly, we make the test mass $m$ unchanged, and the oscillation periods $T_3$ and $T_4$ are measured when the dc voltage is set as $U=U_1$ and $U=U_2$. The calculated $\beta_{2}$ in electrical unit is as
\begin{equation}
\beta_{2}=\frac{k(U_{1}^{2}T_{3}^{2}-U_{2}^{2}T_{4}^{2})}{T_{3}^{2}-T_{4}^{2}}.
\end{equation}
Then the SI-1990 electrical ratio $\gamma$ and the Planck constant $h$ are respectively determined as \cite{A1}
\begin{equation}
\gamma=\frac{\{\frac{4\pi^{2}}{T^{2}}(\beta_{1}+2m)\}_\mathrm{SI}}{\{\beta_{2}-kU^{2}\}_{90}},
\end{equation}
\begin{equation}
h=\frac{4\gamma}{R_{K-90}K_{J-90}^{2}},
\end{equation}
where $R_{K-90}$ and $K_{J-90}$ are conventional values for the von Klitzing constant and the Josephson constant \cite{9}. The measurement for the quasi-elastic electrostatic oscillation method is divided into two separated phases. One is to measure the capacitance coefficient $k$ by measuring $\Sigma C=C_{12A}+C_{12B}+C_{13A}+C_{13B}$ as a function of the vertical displacement $z$. The other phase is measuring oscillation periods $T_1$, $T_2$, $T_3$ and $T_4$ at different conditions.

It can be seen all measurement quantities: the displacement of electrodes 2, the capacitance of $C_{12}$ and $C_{13}$, the applied dc voltage, and the oscillation period, in theory, can be measured accurately. Besides, it is noticed that three approximations are applied in the approach: 1) $m_1=m_2=m$, 2) $L_1=L_2=L$, and 3) $\theta\rightarrow 0$. Approximations 1) and 2) are obtained by adjusting and exchanging two equal masses in left and right weighing pans. Approximation 3) can be corrected by linear extrapolations at small measurement intervals.

\section{Apparatus}
\subsection{Overview}
A prototype of the inertial mass project has been built as shown in Fig. 2. A conventional beam balance with 0.5m beam length is employed as the mainstay for modifications. The mass center of the balance beam is adjusted to a low position below the central knife by taking off the aluminum block above the central knife and adding two adjustable copper blocks below the central knife. A magnetic velocity sensor is designed for compensating the energy consumption during the oscillation by passing a current into a linear actuator. The weighing pans are connected to side knives by flexible structures. A servo system is used to take test masses (1kg, 2kg) synchronously on and off weighing pans. Two Kelvin capacitors (copper, gold-coated) with the same geometry parameters are symmetrically assembled below the weighing pans, and the distance between movable electrode 2 and fixed electrodes 1 (3) is 10mm at balancing position. A laser beam of an interferometer is set in the left pan for measuring positions of movable electrodes. The whole apparatus is placed on an isolated platform with reduced ground vibrations. A glass chamber is used to cut off the air flow. The room temperature is controlled with $\pm0.5^{\circ}$C by air conditioning system.
\begin{figure}[!t]
\centering
\includegraphics[width=0.55\columnwidth]{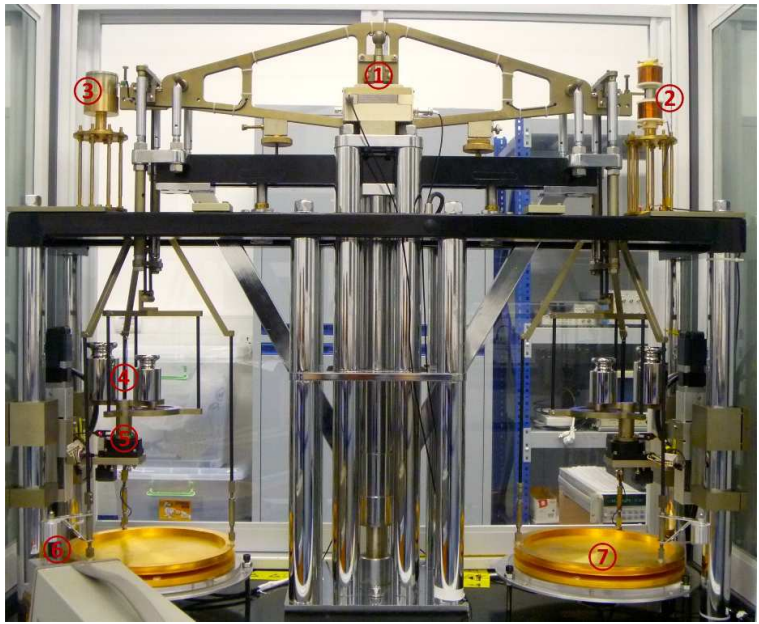}
\caption{The prototype of the inertial mass measurement project. \ding{172} balance beam (central knife); \ding{173}  actuator; \ding{174}  velocity sensor; \ding{175}  test mass; \ding{176}  mass servo system; \ding{177}  laser beam of an interferometer; \ding{178}  Kelvin capacitor. }
\label{fig_2}
\end{figure}

\subsection{Velocity sensor and actuator}
\begin{figure}[!t]
\centering
\includegraphics[width=0.55\columnwidth]{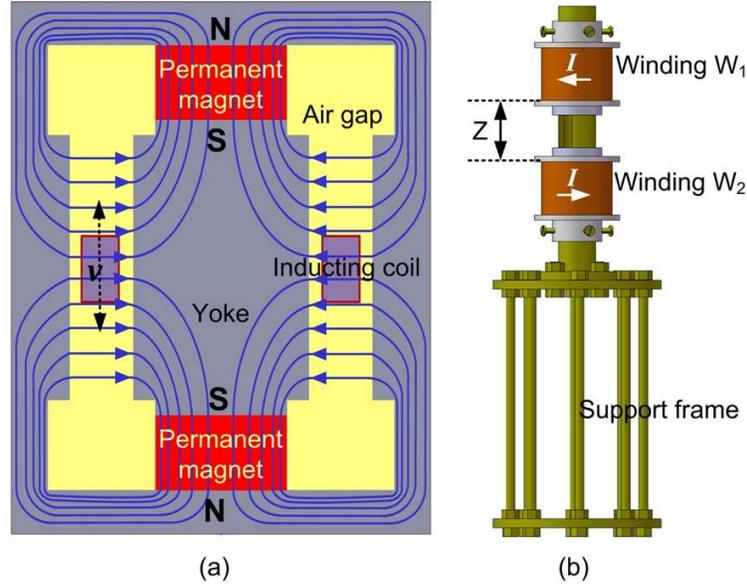}
\caption{Schematic diagrams of (a) velocity sensor and (b) actuator. }
\label{fig_3}
\end{figure}
A magnetic velocity sensor and an linear actuator are employed. The velocity sensor is a 20000-turn copper coil fixed at the left end of the balance beam. The magnetic circuit is shown in Fig.3 (a). Two opposite faced permanent magnets are set in the iron magnet yoke to supply a radial magnetic flux density in the air gap with approximate uniformity in the measurement range of $\pm$2mm along the vertical direction. The output signal is the induced voltage when the coil is moving in the air gap. Note that the coil frame is made by glass material to avoid eddy currents during the oscillation.

The schematic diagram of the actuator is shown in Fig.3 (b). Two windings $W_{1}$ and $W_{2}$ are connected in subtractive series and excited by the same dc current. A permanent magnet is fixed in the middle of $W_{1}$ and $W_{2}$ at the right end of the balance beam to drive the beam into different positions by exciting different amplitude dc currents in windings. The linearity of the actuator is adjusted by changing the vertical distance $Z$ of two windings. The measured function between the vertical displacement $z$ of the movable electrode and the excitation current in windings when $Z=40$mm is as shown in Fig.4. In the range of $\pm$2mm, the actuator has a linearity of 0.5\%, and the residual value performs as a cubic function with the excitation current.
\begin{figure}[!t]
\centering
\includegraphics[width=0.6\columnwidth]{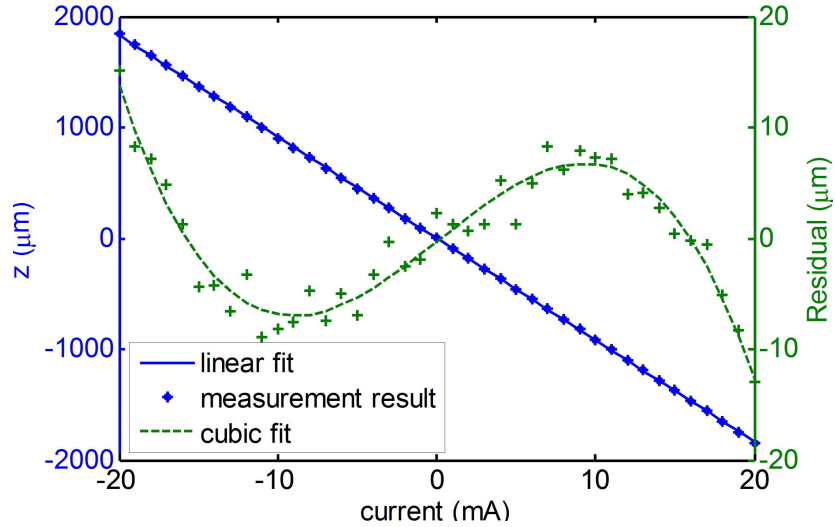}
\caption{Function between the movable electrode position $z$ and the excitation current in actuator. $(*)$ is the measured value; $(+)$ is the residual value.}
\label{fig_4}
\end{figure}

\subsection{Weighing sensitivity and beam length equality}
The mass center modification of balance beam will reduce the weighing sensitivity for the beam balance in theory. However, the high resolution of capacitance measurement for the Kelvin capacitor makes up, or even improves the weighing sensitivity. The test result of $C_{12A}$ with weighing different small values masses is shown in Fig.5. The initial state for the balance was weighing two 1kg test masses. When sheet masses (50mg, 100mg) were added on or taken off weighing pans, the capacitance $C_{12A}$ changed obviously with a sensitivity of 0.012pF/mg. Compared to the sensitivity of 1mg before mass center modification, the weighing sensitivity for the balance is now improved to several tens of $\mu$g.

In the approach, test masses and beam lengths should be equal, i.e., $m_{1}=m_{2}=m$ and $L_{1}=L_{2}=L$. The equivalence of test masses can be easily realized by precision definition. The beam length equality is adjusted by exchanging test masses in pans to make a torque balance, i.e.,$m_1gL_1=m_2gL_2$, $m_2gL_1=m_1gL_2$. $E_2$ class masses are used for preliminary tests. The capacitance $C_{12A}$ with weighing different mass is shown in Fig.5. It can be seen when 1kg test masses are take off two pans ($m_{1}=m_{2}=0$kg) or two 2kg masses are added ($m_{1}=m_{2}=2$kg), the capacitance change of $C_{12A}$ is less than 0.01pF, therefore the beam length equality is about several parts in $10^{7}$.
\begin{figure}[!t]
\centering
\includegraphics[width=0.6\columnwidth]{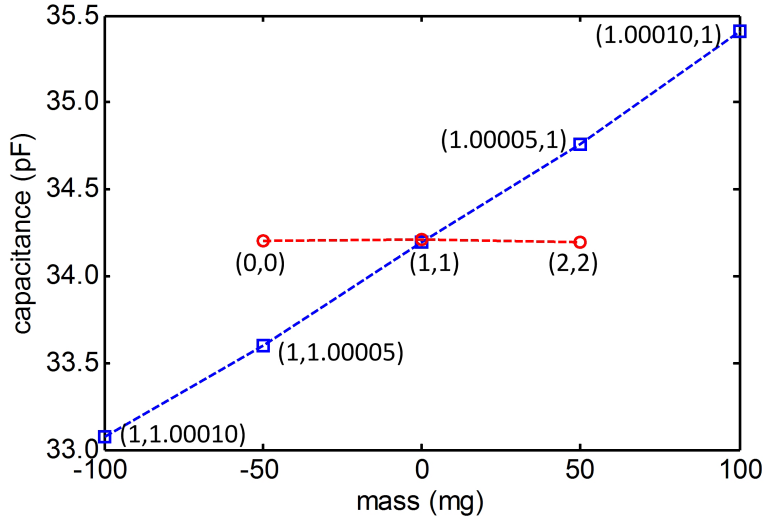}
\caption{Test results of weighing sensitivity and beam length equality for the beam balance. The marked horizontal coordinate is the mass weighing in left pan while the vertical coordinate is the mass in right pan (unit, kg).}
\label{fig}
\end{figure}

\subsection{Voltage source}
\begin{figure}[!t]
\centering
\includegraphics[width=0.55\columnwidth]{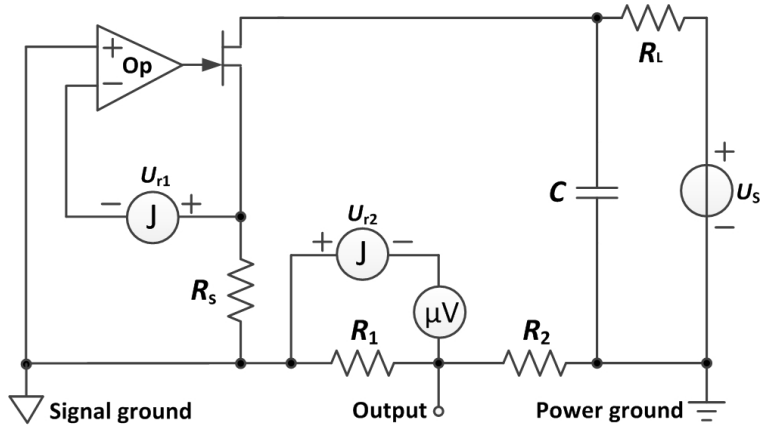}
\caption{Circuit of the designed voltage source. Several terminals are placed in $R_{2}$ to output different dc voltages (0V-700V-1000V-1200V-1400V-1600V). }
\label{Voltage_Circuit}
\end{figure}
\begin{figure}[!t]
\centering
\includegraphics[width=0.6\columnwidth]{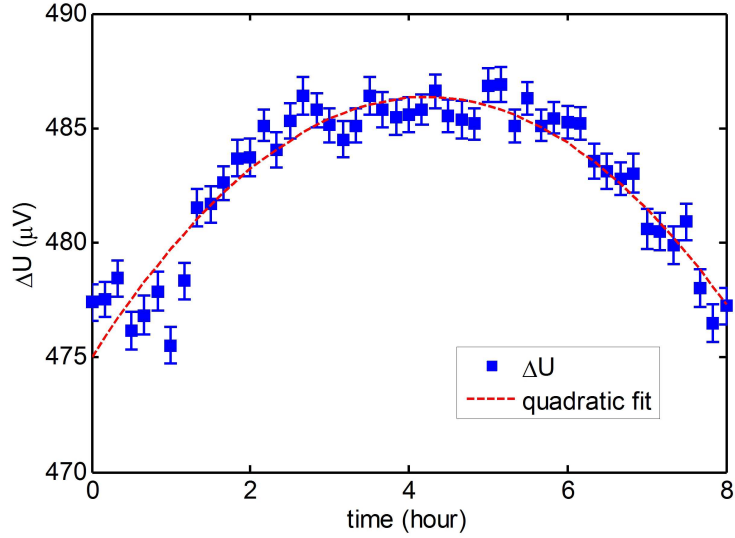}
\caption{A typical stability test of the voltage source. Each measurement contains 300 points. Note the quadratic drift repeats with several $\mu$V.}
\label{Voltage_Output}
\end{figure}

The dc voltage source is designed as a negative feedback system based on a 160:1 resistance divider as shown in Fig.\ref{Voltage_Circuit}. $U_{S}$ is a adjustable rippled dc voltage supply from a rectifier (up to 3000V). A power resistor $R_L$ (100$\Omega$, 100w) is to limit the charge current and capacitor $C$ (5000$\mu$F) is used as a smoothing filter. $U_{r1}$ and $U_{r2}$ are both 10V voltage references (Fluke 732B) and $R_{s}=100$k$\Omega$ is a sampling resistor with temperature coefficient lower than 1ppm/$^{\circ}$C. By the feedback of the regulator, a 100$\mu$A dc current is passing through the resistance divider $R_{1}=100$k$\Omega$ (100k$\Omega\times$1) and $R_{2}$=16M$\Omega$ (100k$\Omega\times$160). In order to obtain a stable resistance ratio, all elements in $R_{1}$ and $R_{2}$ are the same type resistors with similar temperature coefficient lower than 1ppm/$^{\circ}$C. A typical experimental test in Fig.\ref{Voltage_Output} shows the stability of the designed voltage source is about 1ppm (peak-peak) in 8 hours. Note the test is operated in air and the quadratic drift is caused due the self heating of the resistance divider, which is considered to be improved by a better temperature control system (e.g., a oil tank) in future.

\section{Experiment and Discussion}
\subsection{Capacitance coefficient measurement}
\begin{figure}[!t]
\centering
\includegraphics[width=0.65\columnwidth]{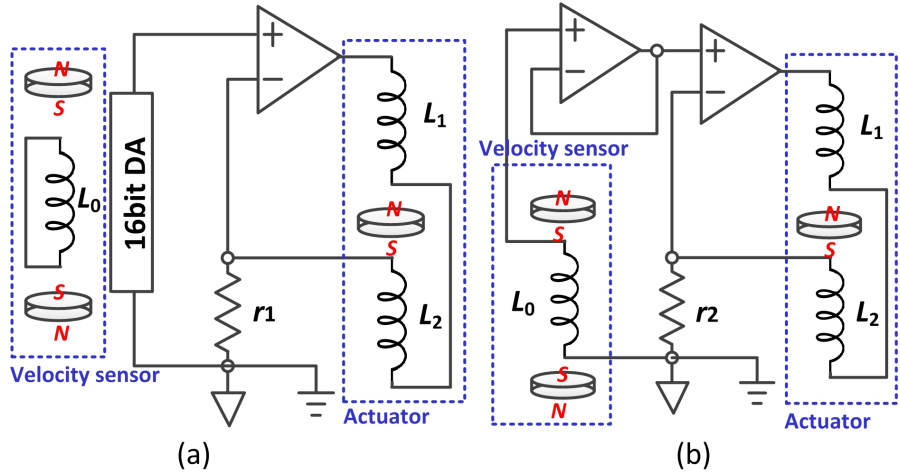}
\caption{Controlling circuits at different measurement phases. (a) capacitance coefficient measurement; (b) periods measurement.}
\label{Control}
\end{figure}

Two capacitances for each Kelvin capacitor, $C_{12}$ and $C_{13}$, are measured by a commercial transformer capacitance bridge AH2700A at different suspension positions. The positioning is currently controlled by an open-loop circuit as shown in Fig.\ref{Control}~(a). The 20bit DA outputs dc voltage between -5V and +5V. Resistor $r_{1}=100\Omega$ is made by twenty 500$\Omega$ elements (2 series as one component, 10 components in parallel) to reduce the heating problem. The velocity sensor is shorted as a damper in capacitance measurement phase to reduce unwanted mechanical vibrations. The position of movable electrodes is measured by an interferometer.

\begin{figure}[!t]
\centering
\includegraphics[width=0.6\columnwidth]{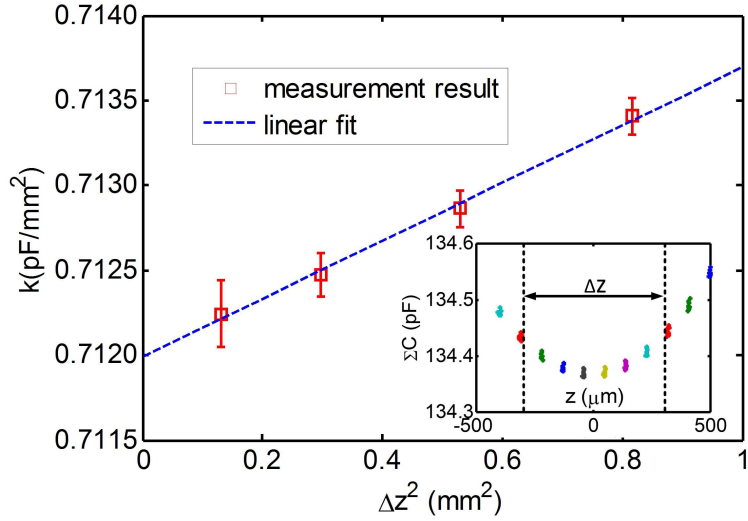}
\caption{Experimental results for measuring the capacitance coefficient $k$.}
\label{k}
\end{figure}

To reduce the uncertainty from mechanical vibrations, the capacitance and the vertical position are synchronously measured. Besides, as the mathematical models of $C_{12}$ and $C_{13}$ are well known \cite{12}, a best fit according to the models is applied in the data analysis. A typical measurement result of function $\Sigma C(z)$ in range of $-500\mu$m$<z<500\mu$m is shown in Fig.\ref{k}. Based on the symmetry of the twin Kelvin capacitor system, the function between $\Sigma C(z)$ and $z$ should follow the following equation
\begin{equation}
\Sigma C(z)=C_{0}+\alpha_{2}z^{2}+\alpha_{4}z^{4}+...,
\end{equation}
where $C_{0}$ is the fixed component; $\alpha_{2}$ and $\alpha_{4}$ are the second-order and fourth order coefficients.
As the oscillation equation is obtained with zero amplitude, the nonlinearity of the measurement must be corrected. Here a linear extrapolation between the calculated $k$ and the length of the fit interval $\Delta z$ (shown in Fig.\ref{k}) is applied.
And $k$ is determined as the value when $\Delta z=0$ as
\begin{equation}
k=\frac{\partial^{2}\Sigma C(z)}{\partial z^{2}}|_{\Delta z=0}.
\end{equation}

The linear extrapolation result is shown in Fig.\ref{k}. The $k$ value with zero oscillation amplitude is 7.11966pF/mm$^{2}$. The error bars (standard deviation) are about several parts in $10^{4}$ and the residual values of the linear fit are several parts in $10^{5}$.

\subsection{Periods measurement}
It is known that the damping of the system, which is mainly caused by air resistance and mechanical friction, will slow down the oscillation. In the periods measurement phase, a linear velocity feedback circuit is designed as shown in Fig.\ref{Control}(b) to compensate the energy loss in each period. The compensation can be expressed mathematically as the following equation
\begin{equation}
(\beta_{1}+2m)\frac{d^{2}\theta}{dt^{2}}+(\xi-\frac{\varepsilon_{1}\varepsilon_{2}}{r_{2}})\frac{d\theta}{dt}+(\beta_{2}-kU^{2})\theta=0,
\end{equation}
where $\xi$ is the natural damping ratio. $\varepsilon_1$ is the proportion of the velocity sensor as $u=\varepsilon_1d\theta/dt$ ($u$ is the output of the velocity sensor) while $\varepsilon_2$ is the proportion of the actuator as $\tau_0=\varepsilon_2u/r_{2}$ ($\tau_{0}$ is the output moment of the actuator). It can be seen that the damping will be changed by choosing different resistance values of $r_2$.

\begin{figure}[!t]
\centering
\includegraphics[width=0.6\columnwidth]{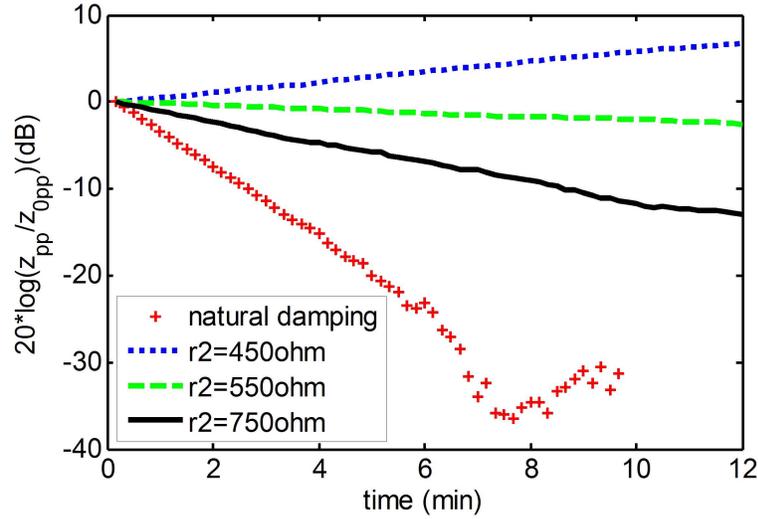}
\caption{Damping behaviors of the quasi-elastic electrostatic oscillation system. $z_{pp}$ is the peak-peak value of the oscillation amplitude and $z_{0pp}=1700\mu$m is its initial value.}
\label{Damping}
\end{figure}
Experimental damping behaviors of the oscillation system is shown in Fig.\ref{Damping}. It can be seen that the natural damping ratio $\xi$ is approximate constant when the oscillation amplitude $z_{pp}>200\mu$m. In the measurement, the resistor $r_2$ is selected with resistance value of 550$\Omega$ to keep the attenuation of the oscillation in a slow speed of 0.2dB/min. During the oscillation, both the oscillation period and the amplitude are simultaneously measured. The period is measured using a commercial frequency counter SR620 triggered by a rectified square waveform signal, which is converted from the velocity signal by a rectifier.

Note that for a under damping system (the damping coefficient $0<\varsigma<1$), the damping will introduce an error $e_d$ for the period measurement, expressed as
\begin{equation}
e_d=\frac{\omega-\omega_0}{\omega_0}=\sqrt{1-\varsigma^2}-1\approx-\frac{\varsigma^2}{2},
\label{eq.ed}
\end{equation}
where $\omega$ is the damped frequency and $\omega_0$ is the frequency without any damping. In the presented case when $r_2=550\Omega$, the damping coefficient $\varsigma$ is calculated as $6\times10^{-4}$ with a typical oscillation period of 10 seconds, and hence the error for measuring the oscillation period due to the damping effect is $-1.8\times10^{-7}$ according to (\ref{eq.ed}). To achieve the measurement uncertainty of 2 parts in $10^{8}$ for the Planck constant determination, $e_d$ should be corrected with at least a 0.1 accuracy level.

Similar to the capacitance coefficient measurement, the nonlinear correction for periods measurement is also required. As restoring moments, either the mechanical component or the electrostatic component, are odd functions, thus the oscillation period $T$ performs as an even function as
\begin{equation}
T=T_{0}(1+\rho_{2}z_{0}^{2}+\rho_{4}z_{0}^{4}+...),
\end{equation}
where $T_{0}$ is the oscillation period with zero amplitude; $z_{0}$ is the oscillation amplitude; $\rho_{2}$ and $\rho_{4}$ are Taylor coefficients.

\begin{figure}[!t]
\centering
\includegraphics[width=0.55\columnwidth]{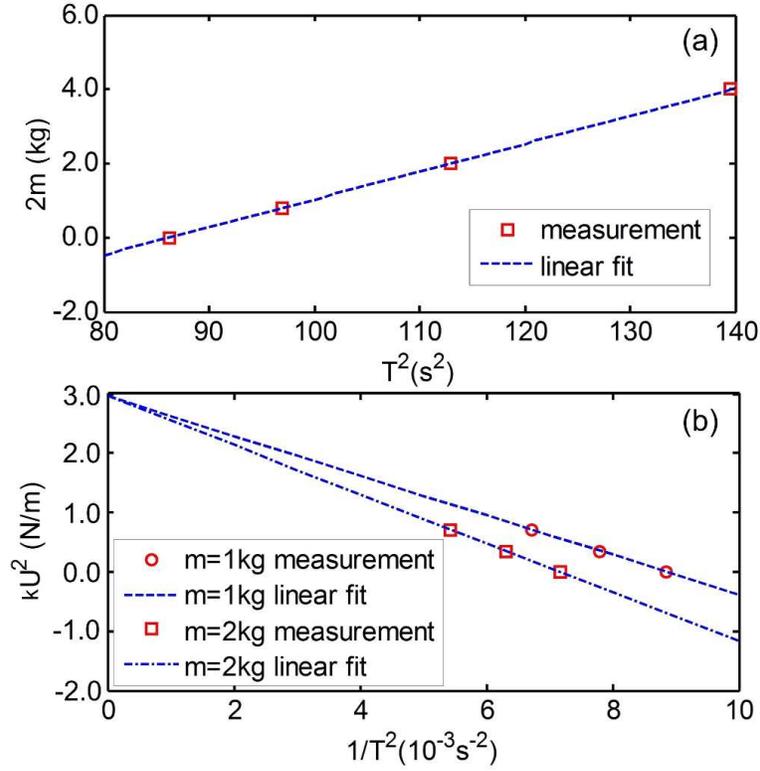}
\caption{Measurement relations of periods with (a) different test masses (0kg, 0.4kg, 1kg, 2kg) and (b) different voltages (0V, 700V, 1000V).}
\label{TT}
\end{figure}

\begin{figure}[!t]
\centering
\includegraphics[width=0.6\columnwidth]{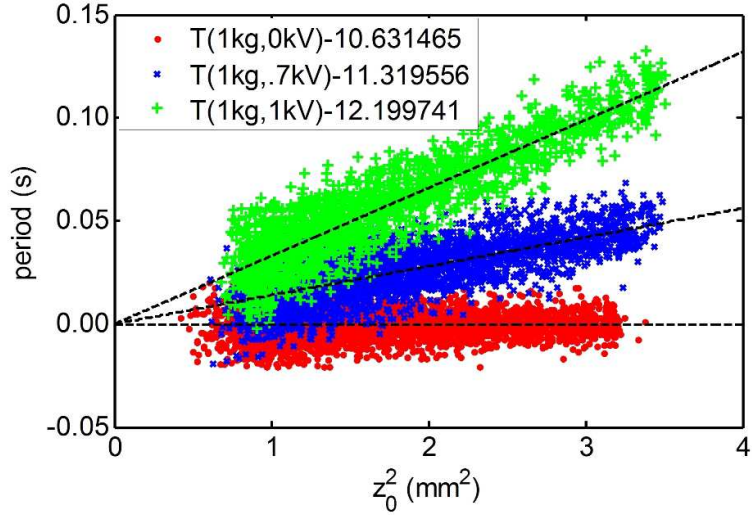}
\caption{A typical nonlinearity measurement of $T$ and $z_0^2$ when $m=1$kg. The dashed line ($--$) is the linear fit for each measurement.}
\label{T1}
\end{figure}

The measurement relations of periods with different test masses (0kg, 0.4kg, 1kg, 2kg) and different voltages (0V, 700V, 1000V) are shown in Fig.\ref{TT} \cite{13}. Note all the period values applied are with zero amplitude, which are calculated by linear extrapolations of $T$ and $z_{0}^2$. A typical nonlinearity measurement of $T$ and $z_0^2$ when $m=1$kg is demonstrated in Fig.\ref{T1}. For each period measurement, the standard deviation is about several parts in $10^{5}$. It is concluded from Fig.\ref{T1} that the electrostatic restoring moment performs a stronger nonlinearity than the mechanical component, which has been discussed in \cite{14}.

\subsection{The Planck constant}
 Based on the measured oscillation period $T_0$ as a two-dimensional function of the weighing mass $m$ and the applied dc voltage $U$, both the SI value and 1990 conventional electrical value for both $\beta_1$ and $\beta_2$ can be calculated by a least-squares fit. Knowing the SI-1990 electrical ratio $\gamma$, the Planck value is obtained by (7). In the analysis, no significant systematic error is found on the principle demonstration for measuring the Planck constant with a relative uncertainty of several parts in $10^4$.

 It is found by experiment that the main uncertainty (3 parts in $10^{4}$) of the measurement is caused by the mechanical deformation of the balance beam when different test masses are added on the mass pan. On the current stage, the balance beam is simply realized by modification of a conventional weighing balance, whose rigidity, however, is not strong enough to ensure the stability of $\beta_2$ with different masses. A further wheel realization of the balance beam, which is similar to the NIST-3 watt balance design \cite{15} with optimized moment of inertia, can reduce the deformation effect by a factor of more than 20. In the meanwhile, an accurate correction model of this effect based on limited measurements are under development. We hope the related uncertainty component can be suppressed below $1\times10^{7}$ by conjunction with mechanical optimizations and corrections.

\section{Conclusion}
A inertial mass measurement approach, the quasi-elastic electrostatic oscillation project, is introduced at NIM. The method avoids the difficult mass center measurement of a conventional inertial mass determination. The absolute and 1990 electrical units are related by a torque transformation on a beam balance oscillator using twin-Kelvin capacitor system. The principle, a prototype, and several experimental investigations for the inertial mass measurement are presented as the principle demonstration. A relative uncertainty of several parts in $10^4$ for the Planck constant measurement is obtained.

A new measurement system, including the wheel balance beam oscillator, a more precise capacitor manufacture, new sensors, a precision PID position control system, and a whole interferometer system, is now being considered to reduce the measurement uncertainty. A result comparison between the inertial mass measurement and the gravitational mass measurement of the Joule Balance experiment is as a first step expected in the future.

\section*{Acknowledgment}

The authors would like to thank Mr. Nong Wang and Mr. Jinxin Xu for valuable discussions of designing the dc voltage source. This project is supported by the National Natural Science Foundation of China (Grant No. 51477160) and the National Department Public Benefit Research Foundation (Grant No. 201010010).

\end{document}